\documentclass{ws-jnmp}
\usepackage{graphicx}
\usepackage{dcolumn}
\usepackage{bm}
\begin{document}
\catchline{}{}{2012}{}{}
\title{Method of generating $N$-dimensional isochronous nonsingular Hamiltonian systems}

\author{A. Durga Devi}
\address{Centre for Nonlinear Dynamics, \\School of Physics, Bharathidasan University,\\ Tiruchirappalli - 620 024, India.}
\author{ R. Gladwin Pradeep}
\address{Centre for Nonlinear Dynamics, \\School of Physics, Bharathidasan University,\\ Tiruchirappalli - 620 024, India.}
\author{V. K. Chandrasekar}  
\address{Centre for Nonlinear Dynamics, \\School of Physics, Bharathidasan University,\\ Tiruchirappalli - 620 024, India.}
\author{M. Lakshmanan}
\address{Centre for Nonlinear Dynamics, \\School of Physics, Bharathidasan University,\\ Tiruchirappalli - 620 024, India.\\
\email{lakshman@cnld.bdu.ac.in}}

\date{\today}
\maketitle
\begin{abstract}
  In this paper we develop a straightforward procedure to construct higher dimensional isochronous Hamiltonian systems.  We first show that a class of singular Hamiltonian systems obtained through the $\Omega$-modified procedure is equivalent to constrained Newtonian systems.  Even though such systems admit isochronous oscillations, they are effectively one degree of freedom systems due to the constraints.  Then we generalize the procedure in terms of $\Omega_i$-modified Hamiltonians and identify suitable canonically conjugate coordinates such that the constructed $\Omega_i$-modified Hamiltonian is \emph{nonsingular} and the corresponding Newton's equation of motion is constraint free.  The procedure is first illustrated for two dimensional systems and subsequently extended to $N$-dimensional systems.  The general solution of these systems are obtained by integrating the underlying equations and is shown to admit isochronous as well as amplitude independent quasiperiodic solutions depending on the choice of parameters.  

\end{abstract}

\section{Introduction}
Nonlinear oscillators model many physical phenomena and the study of such oscillators plays an important role in the theory of nonlinear dynamical systems.  Various methods  have been developed to study nonlinear oscillator equations.  The dependence of frequency on the amplitude of oscillation is generally considered to be a fundamental property of a nonlinear oscillator.  However, recently several nonlinear oscillators have been identified whose frequencies are constants independent of the initial condition, and are known as isochronous systems.  Moreover, several procedures have been developed to construct and identify classes of isochronous oscillators.  In particular Calogero \cite{Calogero:08,Calogero:08f} and Calogero and Leyvraz \cite{Calogero:08g,Calogero:08d,Calogero:08c,Calogero:07,Calogero:08a} have developed many techniques to generate isochronous oscillators and one among these is the method of transforming a real autonomous Hamiltonian $H(\underline{p},\underline{q})$ into a $\Omega$-modified Hamiltonian $\tilde{H}=\frac{1}{2}(H(\underline{p},\underline{q})^2+\Omega^2Q(\underline{p},\underline{q})^2)$, where $\underline{p}=(p_{1},p_{2},.....p_{N})$ and  $\underline{q}=(q_{1},q_{2},...q_{N})$.  Note here that $\tilde{H}$ is the Hamiltonian of the one dimensional harmonic oscillator 
$\ddot{Q}+\Omega^2Q=0,$
and $(H,\,Q)$ are the canonical conjugate momentum and coordinate, respectively.  It is interesting to note that the nonlinear equation obtained from this $\Omega$ modified Hamiltonian is isochronous and the system evolves periodically with the fixed period $T=2\pi/\Omega$.

Recently Chandrasekar et al proposed a procedure to generate scalar isochronous systems recursively from a given Hamiltonian \cite{CDL:09}.  In a recent paper, a procedure to identify several classes of coupled isochronous systems has been proposed and these systems are constructed  using certain nonlocal transformations on a system of harmonic oscillators \cite{jmp-glad-2010}.  In the present paper, we briefly analyse the above mentioned procedure of $\Omega$-modified Hamiltonian developed to generate a system of $N$-dimensional isochronous oscillators.  We find that while the systems generated by this procedure may be isochronous, the underlying Hamiltonian turns out to be singular or degenerate \cite{singular,singular2,singular3} as $\displaystyle\vline\frac{\partial^2 \tilde{H}}{\partial p_i\partial p_j}\vline=0$ for this class.  Consequently the Newton's equation of motion becomes constrained with $(N-1)$ holonomic constraints so that the resultant system corresponds effectively to a one degree of freedom system.  Without these constraints, the Newton's equation of motion will admit unbounded solution, while with the introduction of the constraints the solution becomes bounded and isochronous.  Consequently we find that for such singular Hamiltonians the set of solutions of the Hamilton's equation of motion is contained in, but in general not equal to the set of solutions of the corresponding Newton's equation of motion.  Also we note that while all the solutions of the Hamilton's equation of the systems discussed in this paper are  indeed isochronous, the same is not true for the solutions of the Newton's equations. 

In this paper we show how one can develop a procedure which leads to nonsingular Hamiltonian systems such that the Hamilton's equation or Newton's equation of motion are constraint free, but which can still admit isochronous oscillations depending on the choice of parameters.  For other choices one can obtain amplitude independent quasiperiodic solutions.  The procedure is first demonstrated for two-degrees of freedom systems and subsequently extended to $N$-degrees of freedom systems. 

 We organize our study as follows.  In section 2, we briefly discuss the procedure of $\Omega$-modified Hamiltonian developed by Calogero and Leyvraz \cite{Calogero:08a} and note that the deduced Hamiltonian is a singular or degenerate one.  In section 3, we modify this procedure suitably and use it to construct two-degrees of freedom isochronous systems which are free from constraints.  We obtain the general solution and find it to be isochronous or quasiperiodic for appropriate choice of parameters.  In section 4, we extend the procedure to $N$-degrees of freedom to construct isochronous systems.  Finally, in section 5 we summarize our results.  In the Appendix we point out a procedure to obtain the general solution of certain coupled nonlinear differential equations which are of interest for the present study.

\section{The method of $\Omega$-modified Hamiltonian}    
Let us consider an $N$-dimensional system with a Hamiltonian of the form
\begin{eqnarray} 
H(\underline {p},\underline {q})=\sum_{n=1}^{N}a_{n}p_{n}[\frac{\partial Q(\underline {q})}{\partial q_{n}}]^{-1}, 
\end{eqnarray}
where $\underline{p}=(p_{1},p_{2},...,p_{N})$ and $\underline {q}=(q_{1},q_{2},...,q_{N})$ and $Q(\underline {q})$ is an arbitrary function of the canonical coordinates $q'_{n}s,\,\, n=1,2,...N$. The form of the Hamiltonian is chosen in such a way that the Poisson bracket $\{H,Q\}$ is unity for the condition $\sum_{n=1}^{N}a_{n}=1$. In other words, the Hamiltonian $H(\underline{p},\underline{q})$ and $Q(\underline{q})$ are the canonically conjugate variables for the choice $\sum_{n=1}^{N}a_{n}=1$.  

Using the above canonically conjugate coordinate and momentum variables the procedure of Calogero and Leyvraz \cite{Calogero:08a} allows one to construct the $\Omega$-modified Hamiltonian $\tilde H(\underline {p},\underline {q};\Omega)$ through the relation
\begin{eqnarray}
\tilde H(\underline {p},\underline {q};\Omega)&=&\frac{1}{2}\bigg[\left(H(\underline {p},\underline {q})\right)^2+\Omega^2(Q(\underline {q}))^2\bigg],\\
&=&\frac{1}{2}\left[\left\{\sum_{n=1}^{N}a_{n}p_{n}\left(\frac{\partial Q(\underline {q})}{\partial q_{n}}\right)^{-1}\right\}^2+\Omega^2Q(\underline{q})^2\right],
\label{isoham1}
\end{eqnarray}
where $\Omega$ is a constant.  The above Hamiltonian produces the following $2N$ first order canonical equations of motion for the canonical coordinates $q_n$ and $p_n$,
\begin{subequations}
\label{isoeq1-can}
\begin{eqnarray}
&&\dot{q}_n=a_n\left[\frac{\partial Q(\underline{q})}{\partial q_n}\right]^{-1}H,\qquad n=1,2,\ldots,N\label{isoeq1-can1}\\
&&\dot{p}_n=H\sum_{m=1}^N\left\{a_mp_m\left[\frac{\partial Q(\underline{q})}{\partial q_m}\right]^{-2}\left[\frac{\partial^2Q(\underline{q})}{\partial q_m\partial q_n}\right]\right\}-\Omega^2Q(q)\frac{\partial Q(\underline{q})}{\partial q_n}.\label{isoeq1-can2}
\end{eqnarray}
\end{subequations}
A given Hamiltonian is known as a singular/degenerate Hamiltonian if the determinant of the Hessian matrix is zero, that is the condition
\begin{eqnarray}
\vline\frac{\partial^2\tilde{H}}{\partial p_i\partial p_j}\vline=0,\quad i,j=1,2,\ldots,N,
\end{eqnarray}
is satisfied \cite{singular,singular2,singular3}.  Now substituting the form of the $\Omega$-modified Hamiltonian $\tilde{H}$ given by equation (\ref{isoham1}) into the Hessian matrix, we get
\begin{eqnarray}
\left|\frac{\partial^2\tilde{H}}{\partial p_i\partial p_j}\right|=\left|
\begin{array}{cccc}
a_1^2\left(\frac{\partial Q}{\partial q_1}\right)^{-2}&a_1a_2\left(\frac{\partial Q}{\partial q_2}\right)^{-1}\left(\frac{\partial Q}{\partial q_1}\right)^{-1}&\ldots&a_1a_N\left(\frac{\partial Q}{\partial q_N}\right)^{-1}\left(\frac{\partial Q}{\partial q_1}\right)^{-1}\\
a_1a_2\left(\frac{\partial Q}{\partial q_1}\right)^{-1}\left(\frac{\partial Q}{\partial q_2}\right)^{-1}&a_2^2\left(\frac{\partial Q}{\partial q_2}\right)^{-2}&\ldots&a_2a_N\left(\frac{\partial Q}{\partial q_N}\right)^{-1}\left(\frac{\partial Q}{\partial q_2}\right)^{-1}\\
a_1a_3\left(\frac{\partial Q}{\partial q_1}\right)^{-1}\left(\frac{\partial Q}{\partial q_3}\right)^{-1}&a_2a_3\left(\frac{\partial Q}{\partial q_2}\right)^{-1}\left(\frac{\partial Q}{\partial q_3}\right)^{-1}&\ldots&a_3a_N\left(\frac{\partial Q}{\partial q_3}\right)^{-1}\left(\frac{\partial Q}{\partial q_N}\right)^{-1}\\
\vdots&\ddots&\ddots&\vdots\\
a_1a_N\left(\frac{\partial Q}{\partial q_1}\right)^{-1}\left(\frac{\partial Q}{\partial q_N}\right)^{-1}&a_2a_N\left(\frac{\partial Q}{\partial q_2}\right)^{-1}\left(\frac{\partial Q}{\partial q_n}\right)^{-1}&\ldots&a_N^2\left(\frac{\partial Q}{\partial q_N}\right)^{-2}\\
\end{array}
\right|.
\end{eqnarray}
One can see that taking out the term $a_i\frac{\partial Q}{\partial q_i}$ from each row, where $i$ is the row index, all the rows become equal, that is
\begin{eqnarray}
\left|\frac{\partial^2\tilde{H}}{\partial p_i\partial p_j}\right|=\prod_{i=1}^N\frac{a_i}{\left(\frac{\partial Q}{\partial q_i}\right)}\left|
\begin{array}{cccc}
a_1\left(\frac{\partial Q}{\partial q_1}\right)^{-1}&a_2\left(\frac{\partial Q}{\partial q_2}\right)^{-1}&\ldots&a_N\left(\frac{\partial Q}{\partial q_N}\right)^{-1}\\
a_1\left(\frac{\partial Q}{\partial q_1}\right)^{-1}&a_2\left(\frac{\partial Q}{\partial q_2}\right)^{-1}&\ldots&a_N\left(\frac{\partial Q}{\partial q_N}\right)^{-1}\\
a_1\left(\frac{\partial Q}{\partial q_1}\right)^{-1}&a_2\left(\frac{\partial Q}{\partial q_2}\right)^{-1}&\ldots&a_N\left(\frac{\partial Q}{\partial q_N}\right)^{-1}\\
\vdots&\ddots&\ddots&\vdots\\
a_1\left(\frac{\partial Q}{\partial q_1}\right)^{-1}&a_2\left(\frac{\partial Q}{\partial q_2}\right)^{-1}&\ldots&a_N\left(\frac{\partial Q}{\partial q_N}\right)^{-1}\\
\end{array}
\right|=0,
\end{eqnarray}
which implies the Hamiltonian (\ref{isoham1}) is singular/degenerate.  Consequently, the Newton's equation of motion for such a singular Hamiltonian system is accompanied by a system of constraint equations, that is the corresponding Newtonian equation is a constrained system, as we see below.

In Ref. \cite{Calogero:08a}, the Newton's equation of motion is obtained by differentiating the first equation (\ref{isoeq1-can1}) with respect to $`t$' and replacing $\dot{H}=-\Omega^2Q(\underline{q})$. It is given in the form
\begin{eqnarray}
\label{isoeq1}
\frac{\partial Q(\underline{q})}{\partial q_{n}}\ddot q_{n}+\sum_{m=1}^{N}\bigg(\dot q_{n}\dot q_{m}\frac{\partial^{2}Q(\underline{q})}{\partial q_{n}q_{m}}\bigg)+\Omega^{2} a_{n}Q(\underline{q})=0,n=1,2,\ldots,N.
\end{eqnarray}
Equivalently, one can obtain (\ref{isoeq1}) by differentiating (\ref{isoeq1-can1}) with respect to $t$ and using (\ref{isoeq1-can2}) for $\dot{p}_n$ in it.

However, one can easily check that not all the coordinates $q_i,\,\,i=1,2,\ldots,N$ are independent: there are $(N-1)$ holonomic constraints existing between them.  This can be easily seen by the fact that from (\ref{isoeq1-can1}) it follows that
\begin{eqnarray}
\frac{1}{a_n}\left(\frac{\partial Q}{\partial q_n}\right)\dot{q}_n=H,\quad\mbox{for each}\quad n=1,2,\ldots,N
\end{eqnarray}
Consequently we have the relations
\begin{eqnarray}
\frac{1}{a_1}\dot{q}_1\frac{\partial Q(\underline{q})}{\partial q_1}-\frac{1}{a_j}\dot{q}_j\frac{\partial Q(\underline{q})}{\partial q_j}=0,\quad j=2,3,\ldots,N.\label{constraint1}
\end{eqnarray}
On integration, one obtains a set of $(N-1)$ functional relations on the coordinates $q_i$, $i=1,2,\ldots,N$:
\begin{eqnarray}
\int dq_1\frac{\partial Q}{\partial q_1}-\int dq_j\frac{\partial Q}{\partial q_j}=C_j,\quad j=2,3,\ldots,N,\label{int-constraint}
\end{eqnarray}
where $C_j$'s are constants.  Equation (\ref{int-constraint}) obviously constitutes a set of $(N-1)$ holonomic constraints on the coordinates $q_i$.
These constraints have to be supplemented with the equation of motion (\ref{isoeq1}) which together constitute the Newton's equation of motion.  The resulting system is then equivalent to the Hamilton's equations (\ref{isoeq1-can}).  One may also note that equation (\ref{isoeq1}) on its own merit, that is without the constraints (\ref{int-constraint}), corresponds to the \emph{nonsingular} Hamiltonian
\begin{eqnarray}
\tilde{H}=\frac{1}{2}\left(\sum_{i=1}^Na_i^2p_i^2\left[\frac{\partial Q(\underline{q})}
{\partial q_i}\right]^{-2}+\Omega^2Q(\underline{q})^2\right),
\end{eqnarray}
with the associated canonical equations
\begin{subequations}
\begin{eqnarray}
&&\dot{q}_i=a_i^2p_i\left[\frac{\partial Q(\underline{q})}{\partial q_i}\right]^{-2},\\
&&\dot{p}_i=\sum_{j=1}^Na_j^2p_j^2\left[\frac{\partial Q(\underline{q})}{\partial q_j}\right]^{-3}\frac{\partial^2 Q(\underline{q})}{\partial q_j \partial q_i}-\Omega^2Q(\underline{q})\frac{\partial Q}{\partial q_i}.
\end{eqnarray}
\end{subequations}
\subsection{Specific Examples}
In order to illustrate the above points clearly, we consider a specific choice of $Q(\underline{q})$ which has been considered in Ref. \cite{Calogero:08a}, namely
\begin{eqnarray}
Q(\underline{q})=\sum_{m=1}^N b_mq_m^{k_m},\label{calogero-form}
\end{eqnarray}
where $b_m$'s are arbitrary real parameters and $k_m$'s are such that $(1/k_m)$'s are positive integers. (When $k_m$'s are positive integers, the general solution becomes multivalued/complex solution which we will avoid in our choice of $k_m$'s).  Correspondingly the Newton's equation of motion (without the constraints) (\ref{isoeq1}) takes the form
\begin{eqnarray}
k_nb_n\left[q_n^{k_n-1}\ddot{q}_n+(k_n-1)q_n^{k_n-2}\dot{q}_n^2\right]+\Omega^2a_n\sum_{m=1}^N b_mq_m^{k_m}=0.\label{calogero-newton}
\end{eqnarray}
The associated $(N-1)$ holonomic constraint conditions (\ref{int-constraint}) become
\begin{eqnarray}
\frac{b_1q_1^{k_1+1}}{(k_1+1)}-\frac{b_jq_j^{k_j+1}}{(k_j+1)}=C_j,\quad j=2,3,\ldots,N.\label{gen-constraint}
\end{eqnarray}
The singular Hamiltonian (\ref{isoham1}) for the choice (\ref{calogero-form}) becomes
\begin{eqnarray}
\tilde{H}=\frac{1}{2}\left\{\left(\sum_{n=1}^N\frac{a_np_n}{b_nq_n^{k_n}}\right)^2+\Omega^2\left(\sum_{m=1}^Nb_mq_m^{k_m}\right)^2\right\},\quad \sum_{n=1}^Na_n=1\label{calogero-ham}
\end{eqnarray}
and the canonical equations become
\begin{subequations}
\label{calogero-eq1}
\begin{eqnarray}
&&\dot{q}_i=\frac{a_i}{b_iq_i^{k_i}}\left(\sum_{n=1}^N\frac{a_np_n}{b_nq_n^{k_n}}\right)^2,\\
&&\dot{p}_i=\frac{k_ia_ip_i}{b_iq_i^{k_i+1}}\left(\sum_{n=1}^N\frac{a_np_n}{b_nq_n^{k_n}}\right)-\Omega^2b_ik_iq_i^{k_i-1}\sum_{m=1}^Nb_mq_m^{k_m},\quad i=1,2,\ldots,N.
\end{eqnarray}
\end{subequations}
On the other hand, the nonsingular Hamiltonian corresponding to the Newton's equation of motion (\ref{calogero-newton}) without the constraints (\ref{gen-constraint}) is
\begin{eqnarray}
\tilde{H}=\frac{1}{2}\left\{\sum_{n=1}^N\left(\frac{a_np_n}{b_nq_n^{k_n}}\right)^2+\Omega^2\left(\sum_{m=1}^Nb_mq_m^{k_m}\right)^2\right\},\label{nonsingular-ham}
\end{eqnarray}
with the corresponding Hamilton's equations
\begin{subequations}
\begin{eqnarray}
&&\dot{q}_i=\frac{a_i^2p_i}{b_i^2q_i^{2k_i}},\\
&&\dot{p}_i=\frac{k_ia_i^2p_i^2}{b_i^2q_i^{2k_i+1}}-\Omega^2k_ib_iq_i^{k_i-1}\sum_{m=1}^Nb_mq_m^{k_m},\quad i=1,2,\ldots,N.
\end{eqnarray}
\end{subequations}
We now demonstrate the isochronous nature of the singular Hamiltonian system (\ref{calogero-eq1}) or equivalently the Newton's equation with constraints (\ref{calogero-newton})-(\ref{gen-constraint}), and the nonisochronous and unstable nature of the solutions for the nonsingular Hamiltonian system (\ref{nonsingular-ham}) or equivalently the Newton's equation without the constraints (\ref{calogero-newton}).  To be clear we first discuss the $N=2$ degrees of freedom special case and then the $N$ arbitrary case with a simple choice of parameters.

\subsubsection{$N=2$ case}

Let us consider the specific two degrees of freedom Hamiltonian $\displaystyle H=p_1\sqrt{q_{1}}+p_2\sqrt{q_{2}}$ which is canonically conjugate to the collective coordinate $Q=(\sqrt{q_{1}}+\sqrt{q_{2}})$, that is the Poisson bracket \{H,Q\}=1.  Here we have chosen the constant parameters $a_1=a_2=1/2$, $k_1=k_2=1/2$ in equation (\ref{calogero-form}) for simplicity.

Considering $H$ as the momentum and $Q$ as the canonically conjugate collective coordinate and substituting them into the $\Omega$-modified Hamiltonian (\ref{isoham1}), we get
 \begin{eqnarray}
\tilde H=\frac{1}{2}[H^{2}+\Omega^{2}Q^{2}]=\frac{1}{2}[(p_1\sqrt{q_{1}}+p_2\sqrt{q_{2}})^{2}+\Omega^{2}(\sqrt{q_{1}}+\sqrt{q_{2}})^{2}].\label{iso-cal-ham}
\end{eqnarray}
The corresponding canonical equations of motion are
\begin{subequations}
\label{q1q2-gen}
\begin{eqnarray}
&&\dot{q}_1=H\sqrt{q_1},\qquad
\dot{q}_2=H\sqrt{q_2},\label{q1q2}\\
&&\dot{p}_1=\frac{-1}{2\sqrt{q_1}}\left(p_1H+\Omega^2Q\right),\label{p1}\\
&&\dot{p}_2=\frac{-1}{2\sqrt{q_2}}\left(p_2H+\Omega^2Q\right),\label{p2}
\end{eqnarray}
\end{subequations}
where $H=p_1\sqrt{q_1}+p_2\sqrt{q_2}$ and $Q=(\sqrt{q_1}+\sqrt{q_2})$.  

From (\ref{q1q2-gen}), we can obtain the system of coupled second order ODEs,
\begin{eqnarray}
\ddot q_{1}=\frac{\dot{q}_1^2}{2q_1}-\Omega^2\sqrt{q_1}(\sqrt{q_{1}}+\sqrt{q_{2}}),  \quad \ddot q_{2}=\frac{\dot{q}_2^2}{2q_2}-\Omega^2\sqrt{q_2}(\sqrt{q_{1}}+\sqrt{q_{2}}).\label{isoeq4}
\end{eqnarray}
The general solution of the above system of second order ODEs is (for details see Appendix A)
\begin{subequations}
\begin{eqnarray}
\addtocounter{equation}{-1}
\label{isoeq5}
\addtocounter{equation}{1}
q_1=\frac{1}{4}\left(2\sqrt{q_1(0)}+\frac{H(0)\sin\Omega t}{\Omega}+Q(0)(\cos\Omega t-1)+C(0)t\right)^{2},\\
q_2=\frac{1}{4}\left(2\sqrt{q_2(0)}+\frac{H(0)\sin\Omega t}{\Omega}+Q(0)(\cos\Omega t-1)-C(0)t\right)^{2},
\end{eqnarray}
\end{subequations}
where $q_{1}(0),q_{2}(0),H(0),C(0)$ are the four constants of integration fixed by the initial conditions and $Q(0)=\sqrt{q_{1}(0)}+\sqrt{q_{2}(0)}$.   Note that this solution is aperiodic. Further, one has to also note that the solution to Eq.(2.22) may  become singular due to possible non-uniqueness which may arise when the dynamical variables $q_1(t)$ and $q_2(t)$  pass through the origin. Further, we note here that even though the general solution (\ref{isoeq5}) satisfies the Newton's equation of motion (\ref{isoeq4}), it does not satisfy the Hamilton's equations of motion (\ref{q1q2-gen}). The reason for this is given below.

From equations (\ref{q1q2}), it also follows that
\begin{eqnarray}
\frac{\dot{q}_1}{\sqrt{{q}_1}}=\frac{\dot{q}_2}{\sqrt{{q}_2}}
\end{eqnarray}
leading to the constraint
\begin{eqnarray}
\sqrt{q_1}-\sqrt{q_2}=I,\label{constraint}
\end{eqnarray}
where $I$ is a constant.

Thus the equivalent Newton's equation of motion corresponding to the Hamiltonian system (\ref{q1q2-gen}) is the equation (\ref{isoeq4}) plus the constraint (\ref{constraint}).
This means that only one of the two variables $q_1$ or $q_2$ is independent.   Subject to the constraint (\ref{constraint}), the general solution of (\ref{isoeq4}) can be written in the form
\begin{subequations}
\begin{eqnarray}
\addtocounter{equation}{-1}
\label{isoeq-5}
\addtocounter{equation}{1}
q_1=\frac{1}{4}\left(2\sqrt{q_1(0)}+\frac{H(0)\sin\Omega t}{\Omega}+Q(0)(\cos\Omega t-1)\right)^{2},\\
q_2=\frac{1}{4}\left(2\sqrt{q_2(0)}+\frac{H(0)\sin\Omega t}{\Omega}+Q(0)(\cos\Omega t-1)\right)^{2},
\end{eqnarray}
\end{subequations}
where $\sqrt{q_1(0)}-\sqrt{q_2(0)}=I$, which is periodic, bounded and isochronic, but only with three integration constants $H(0),q_1(0)$ and $q_2(0)$.  A representative solution is also plotted in Fig.1.  One can also check that the solution (\ref{isoeq-5}) is also the solution of the Hamilton's equations (\ref{q1q2-gen}).

Further, we also note that the Newton's equation (\ref{isoeq4}) on its own merit without the constraint (\ref{constraint}) admits a different Hamiltonian of the form
\begin{eqnarray}
H=\left(p_1^2q_1+p_2^2q_2+\frac{\Omega^2}{2}(\sqrt{q_1}+\sqrt{q_2})^2\right),
\end{eqnarray}
which is \emph{not} equivalent to the Hamiltonian (\ref{iso-cal-ham}). The corresponding canonical equations are
\begin{subequations}
\begin{eqnarray}
&&\qquad\qquad\dot{q}_1=2p_1q_1,\qquad \qquad\qquad\dot{q}_2=2p_2q_2,\\
&&\dot{p}_1=-p_1^2-\frac{\Omega^2}{2\sqrt{q_1}}(\sqrt{q_1}+\sqrt{q_2}),\qquad\dot{p}_2=-p_2^2-\frac{\Omega^2}{2\sqrt{q_2}}(\sqrt{q_1}+\sqrt{q_2}),
\end{eqnarray}
\end{subequations}
which is now equivalent to the Newton's equation (\ref{isoeq4}) alone without the constraint (\ref{constraint}). 
\subsubsection{N-arbitrary case}
In the general case, we consider the singular Hamiltonian (\ref{calogero-ham})
whose canonical equations are given by Eqs. (\ref{calogero-eq1}) which lead to $(N-1)$ constraints given by Eq. (\ref{gen-constraint}).  The corresponding Newton's equation of motion is given by Eq. (\ref{calogero-newton}) along with the constraints (\ref{gen-constraint}).
Integrating now the Newton's equation (\ref{calogero-newton}) \emph{without} the constraints (\ref{gen-constraint}), as pointed out in Appendix A, one obtains the general solution as
\begin{eqnarray}
\hspace{-1.8cm}q_n(t)=q_n(0)\left(1+\frac{a_n}{b_n}\frac{1}{(q_n(0))^{k_n}}\left(\frac{H(0)}{\Omega}\sin\Omega t+Q(0)(\cos\Omega t-1)\right)+C_n(0)t\right)^{\frac{1}{k_n}},\label{calogero-gen}
\end{eqnarray}
where $q_n(0)'s$, $\,H(0)$, and $C_n(0)$, $\sum_{n=1}^NC_n=0$, $n=1,2,\ldots,N$ are integration constants fixed by the initial condition and $Q(0)=\sum_{m=1}^Nb_mq_m(0)^{k_m}$.  We note here that the restriction on the parameters $k_m$'s to be reciprocal of real positive integers ensures that the general solution (\ref{calogero-gen}) to be single valued,real and analytic.
Subject to the $(N-1)$ constraints (\ref{gen-constraint}), the Newton's equation admits the $(N+1)$ parameter bounded, isochronous solution
\begin{eqnarray}
&&\hspace{-1cm}q_{n}(t)=q_{n}(0)\bigg(1+\frac{a_{n}}{b_{n}}\frac{1}{(q_{n}(0))^{k_{n}}}\bigg[H(0)\frac{\sin(\Omega t)}{\Omega}+Q(0)(\cos(\Omega t)-1)\bigg]\bigg)^\frac{1}{k_{n}}, \label{isosol1}
\end{eqnarray}
which is also the solution of the Hamilton's equations (\ref{calogero-eq1}).   
Note that the Hamiltonian of the constraint-free equation (\ref{calogero-newton}) is
\begin{eqnarray}
\tilde{H}=\frac{1}{2}\left(\sum_{i=1}^Na_i^2p_i^2\left[\frac{\partial Q(\underline{q})}
{\partial q_i}\right]^{-2}+\Omega^2Q(\underline{q})^2\right).
\end{eqnarray}
Our above analysis clearly shows that for the singular Hamiltonian systems (\ref{isoham1}), the equivalent Newton's equation is a holonomic constrained system (with $(N-1)$ constraint conditions) admitting isochronous oscillatory solution as the general solution.  Consequently, the associated system possesses only one independent coordinate variable.   

Finally, we note that even though the solutions expressed in our above examples lead to isochronous behaviour, they may exhibit in general some kind of singular behaviour either due to non-uniqueness or poles occurring in the solutions depending on the choice of the value of $k_m$'s in (\ref{calogero-form}).

 In the next section, we describe a procedure to modify this system such that the new system with $N$-degrees of freedom admits isochronous oscillations.

\begin{figure}
\centering 
\includegraphics[width=1\columnwidth]{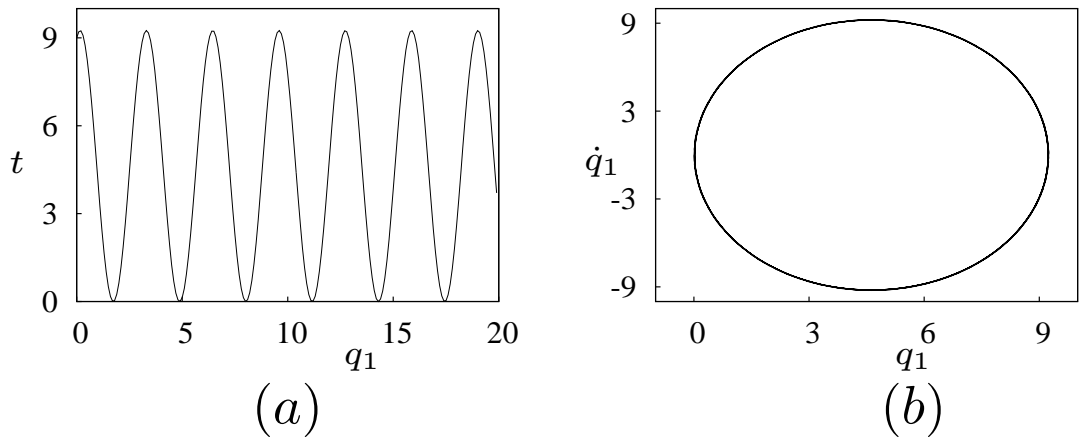}
\caption{\label{fig1} Isochronous oscillation of the system (\ref{isoeq4}) with the constraint (\ref{constraint})  (a) Time series plot for $q_{1}$ (b) Projected $(q_1-\dot{q}_1)$ portrait.}
\end{figure} 


\section{Systematic method to construct higher dimensional isochronous systems}
Even though the procedure of $\Omega$-modified Hamiltonian is an interesting method to construct $N$-dimensional isochronous systems, the $\Omega$-modified Hamiltonian is singular (as seen in the previous section) and the effective degrees of freedom is only one (that is, the number of independent coordinate variables is only one due to the $(N-1)$ holonomic constraints). On inspection for the reason behind this fact, we can easily note that the $\Omega$-modified Hamiltonian is defined in such a way that the original Hamiltonian $H(\underline{p},\underline{q})$, which is a function of $N$ canonical momenta $\underline{p}$ and the $N$ canonical coordinates $\underline{q}$, becomes the canonical momentum of an one dimensional harmonic oscillator.  This mapping of a $N$-dimensional system onto an one dimensional harmonic oscillator is the reason for the inability to rewrite the $2N$ first order ODEs as a system of $N$ coupled second order ODEs without constraints.  One can overcome this problem by suitably redefining the $\Omega$-modified Hamiltonian.  In this section we illustrate this for a class of two-dimensional systems and obtain the general solution for a specific example.  In addition to the periodic isochronous oscillations, the resultant class of two-dimensional systems can also admit amplitude independent quasiperiodic oscillations for appropriate choice of parameters.  

Let us define the $\Omega$-modified Hamiltonian for a two-dimensional system as
\begin{eqnarray}
\tilde{H}=\frac{1}{2}[H_{1}(p_{n},q_{n})^2+H_{2}(p_{n},q_{n})^2+\Omega_{1}^2 Q_{1}(q_{n})^2+\Omega_{2}^2 Q_{2}(q_{n})^2],\,\,n=1,2,\label{isoham2}
\end{eqnarray}
where $H_{1}(p_{n},q_{n})$, $H_{2}(p_{n},q_{n})$ are real Hamiltonians of two arbitrary two-dimensional systems and $Q_{1}$ and $Q_{2}$ are arbitrary functions  canonically conjugate to the Hamiltonians $H_{1}(p_{n},q_{n})$ and $H_{2}(p_{n},q_{n})$, respectively.  We have effectively mapped the Hamiltonians of two different two-dimensional systems to the Hamiltonian of a two-dimensional harmonic oscillator which is now the new $\Omega_{i}$-modified Hamiltonian.  We wish to note here that the new $\Omega_{i}$-modified Hamiltonian $\tilde{H}$ has two natural frequencies of oscillations, $\Omega_{1}$ and $\Omega_{2}$, and depending upon the ratio between them we obtain periodic or quasiperiodic oscillations.

Following the definition of the Hamiltonian $H(\underline{p},\underline{q})$ in the previous section we define the new Hamiltonians $H_1(p_n,q_n)$ and $H_2(p_n,q_n)$ as
\begin{eqnarray}
H_{1}=a_{1}\left(\frac{p_{1}}{Q_{1q_{1}}}\right)+a_{2}\left(\frac{p_{2}}{Q_{1q_{2}}}\right), \quad H_{2}=b_{1}\left(\frac{p_{1}}{Q_{2q_{1}}}\right)+b_{2}\left(\frac{p_{2}}{Q_{2q_{2}}}\right),\label{isoham3}
\end{eqnarray}
where $Q_{iq_j}=\frac{\partial Q_i}{\partial q_j}$, $i,j=1,2$, and $a_i,\,b_i$ are new functions to be determined.    
From the solution of the canonical equations of the two-dimensional harmonic oscillator (\ref{isoham2}) we know that
\begin{eqnarray}
H_{1}=A \cos(\Omega_{1} t+ \delta_{1}),\quad H_{2}=B \cos(\Omega_{2} t+ \delta_{2}),\nonumber \\
Q_{1}=\frac{A}{\Omega_{1}}\sin(\Omega_{1}t+\delta_{1}),\quad Q_{2}=\frac{B}{\Omega_{2}}\sin(\Omega_{2}t+\delta_{2}),\label{harmq}
\end{eqnarray}
where $A$,\,$B$,\,$\delta_{1}$, $\delta_{2}$ are arbitrary constants.

Now, one can easily check that the above Hamiltonians $H_{1}$ and $H_{2}$ are canonically conjugate to the collective coordinates
$Q_1(q_n)$ and $Q_2(q_n)$, respectively, for the choice 
\begin{eqnarray}
a_{1}=\frac{Q_{1q_{1}}Q_{2q_{2}}}{\Delta},\quad a_{2}=-\frac{Q_{2q_{1}}Q_{1q_{2}}}{\Delta},\label{isopar1}\\
b_{1}=-\frac{Q_{1q_{2}}Q_{2q_{1}}}{\Delta},\quad b_{2}=\frac{Q_{1q_{1}}Q_{2q_{2}}}{\Delta},\label{isopar2}
\end{eqnarray}
where $\Delta=Q_{1q_{1}}Q_{2q_{2}}-Q_{1q_{2}}Q_{2q_{1}}\ne0$.  Substituting the above quantities (\ref{isoham3}), (\ref{isopar1}) and (\ref{isopar2}) in the $\Omega$-modified Hamiltonian (\ref{isoham2}) we get
\begin{eqnarray}
\tilde H=&&\frac{1}{2}\bigg[\frac{\big(p_{1} Q_{2q_{2}}-p_{2} Q_{2q_{1}}\big)^2}{\Delta^2}+\frac{\big(p_{2} Q_{1q_{1}}-p_{1} Q_{1q_{2}}\big)^2}{\Delta^2}
+\Omega_{1}^2Q_{1}(q_{1},q_{2})^2+\Omega_{2}^2Q_{2}(q_{1},q_{2})^2\bigg].\label{isoham4}
\end{eqnarray}
Here we note that the above Hamiltonian (\ref{isoham4}) is nonsingular and this can be verified from the relation for the Hessian,
\begin{eqnarray}
\left|
\begin{array}{ccc}
\frac{\partial^2 \tilde{H}}{\partial p_1^2}&\frac{\partial^2 \tilde{H}}{\partial p_1\partial p_2}\\
\frac{\partial^2 \tilde{H}}{\partial p_2\partial p_1}&\frac{\partial^2 \tilde{H}}{\partial p_2^2}
\end{array}
\right|=\Delta
\ne0.
\end{eqnarray}
The choice $\Delta=Q_{1q_{1}}Q_{2q_{2}}-Q_{1q_{2}}Q_{2q_{1}}=0$ implies $Q_1(\underline{q})=f(Q_2(\underline{q}))$, where $f$ is an arbitrary function and for this case the $\Omega$-modified Hamiltonian (\ref{isoham2}) reduces to (\ref{isoham1}), which we have already shown to be a singular one.

The canonical equations of motion corresponding to the above Hamiltonian (\ref{isoham4}) are
\begin{subequations}
\label{2dexample}
\begin{eqnarray}
&&\dot{q}_1=\frac{1}{\Delta^2}\left(Q_{2q_2}(p_1Q_{2q_2}-p_2Q_{2q_1})-Q_{2q_2}(p_2Q_{q_1}-p_1Q_{1q_2})\right),\\
&&\dot{q}_2=\frac{1}{\Delta^2}\left(Q_{1p_2}(p_2Q_{1q_1}-p_1Q_{1q_2})-Q_{2q_1}(p_1Q_{2q_2}-p_2Q_{2q_1})\right),\\
&&\dot{p}_1=\frac{1}{\Delta^3}\bigg[\Delta(p_2Q_{1q_1}-p_1Q_{1q_2})(p_2Q_{1q_{1}q_{1}}-p_1Q_{1q_1q_2})-(p_2Q_{1q_1}-p_1Q_{1q_2})^2\nonumber\\
&&\quad\quad\times\left\{Q_{1q_1}Q_{2q_1q_2}+Q_{2q_2}Q_{1q_1q_1}-Q_{1q_2}Q_{2q_1q_1}-Q_{2q_1}Q_{1q_1q_2}\right\}-(p_1Q_{2q_2}-p_2Q_{2q_1})\nonumber\\
&&\quad\quad\times\Delta(p_1Q_{2q_1q_2}-p_2Q_{q_1q_1})-(p_1Q_{2q_2}-p_2Q_{2q_1})^2\big\{Q_{1q_1}Q_{2q_1q_2}+Q_{2q_2}Q_{1q_1q_1}\nonumber\\
&&\quad\quad-Q_{1q_2}Q_{2q_1q_1}-Q_{2q_1}Q_{1q_1q_2}\big\}\bigg]+\Omega_1^2Q_1Q_{1q_1}+\Omega_2^2Q_2Q_{2q_1},
\end{eqnarray}
and
\begin{eqnarray}
&&\dot{p}_2=\frac{1}{\Delta^3}\bigg[\Delta(p_2Q_{1q_1}-p_1Q_{1q_2})(p_2Q_{1q_1q_2}-p_1Q_{1q_2q_2})-(p_2Q_{1q_1}-p_1Q_{1q_2})^2\nonumber\\
&&\quad\quad\times(Q_{1q_1}Q_{2q_2q_2}+Q_{2q_2}Q_{1q_1q_2}-Q_{1q_2}Q_{2q_1q_2}-Q_{2q_1}Q_{1q_2q_2})-(p_1Q_{2q_2}-p_2Q_{2q_1})\nonumber\\
&&\quad\quad\times\Delta(p_1Q_{2q_2q_2}-p_2Q_{2q_1q_2})-(p_1Q_{2q_2}-p_2Q_{2q_1})^2\big\{Q_{1q_1}Q_{2q_2q_2}+Q_{2q_2}Q_{1q_1q_2}\nonumber\\
&&\quad\quad-Q_{1q_2}Q_{2q_1q_2}-Q_{1q_2q_2}Q_{2q_1}\big\}\bigg]+\Omega_1^2Q_1Q_{1q_2}+\Omega_2^2Q_2Q_{2q_2}.
\end{eqnarray}
\end{subequations}

Rewriting the above Hamilton's equations of motion we get the following system of two coupled second order ODEs,
\begin{eqnarray}
\hspace{-0.5cm}\ddot{q}_i=\frac{1}{\Delta}\left(\sum_{k=1}^2\sum_{j=1}^2A_{ijk} \dot{q}_j\dot{q}_k+B_i\right),\quad i=1,2,\label{two-coupled-eq}
\end{eqnarray}
where
\begin{eqnarray}
\hspace{-2.2cm}A_{ijk}=(-1)^{i+1}\left|\begin{array}{ccc}
Q_{1q_{i+1}}&Q_{2q_{i+1}}\\
Q_{1q_{j}q_k}&Q_{2q_{j}q_k}
\end{array}\right|,
\quad
B_{i}=(-1)^i\left|\begin{array}{cccc}
Q_{1}\Omega_1^2&Q_{2}\Omega_2^2\\
Q_{1q_{i+1}}&Q_{2q_{i+1}}
\end{array}\right|,
\quad Q_j=Q_j(q_1,q_2),2+i=i.
\end{eqnarray}

In order to obtain the explicit general solution of (\ref{2dexample}) or (\ref{two-coupled-eq}) one has to fix the form of $Q_1$ and $Q_2$ in the above equation such that the resultant solutions are analytic and single valued.  For illustration we choose 
\begin{eqnarray}
Q_{1}=k_{1}q_{1}^{r_{1}}+k_{2}q_{2}^{r_{2}}\quad Q_{2}=k_{3}q_{1}^{r_{1}}+k_{4}q_{2}^{r_{2}},\label{harmq1}
\end{eqnarray}
where $r_1$ and $r_2$ are such that $(1/r_1)$ and $(1/r_2)$ are positive integers (see below) so that the resultant solution is single valued and analytic.  One can also choose other forms of $Q_1(q_n)$ and $Q_2(q_n)$ as well but we stick to this form for the sake of simplicity.
Substituting the above forms in (\ref{isoham4}) we get
\begin{eqnarray}
&&\hspace{-2cm}\tilde H=\frac{1}{2} \bigg[\bigg(\frac{(k_{4} p_{1} r_{2} q_1^{1-r_{1}}-k_{3} p_{2} r_{1} q_{2}^{1-r_{2}})}{g_{2}r_{1} r_{2}} \bigg)^{2} +\bigg(\frac{(k_{2} p_{1} r_{2} q_{1}^{1-r_{1}}-k_{1} p_{2} r_{1} q_{2}^{1-r_{2}})}{g_{2}r_{1} r_{2}} \bigg)^{2}\nonumber\\&&+\Omega_{1}^2(k_{1}q_{1}^{r_{1}}+k_{2}q_{2}^{r_{2}})^2+\Omega_{2}^2(k_{3}q_{1}^{r_{1}}+k_{4}q_{2}^{r_{2}})^2\bigg],\label{isoham5}
\end{eqnarray}
where $g_{2}=k_{2}k_{3}-k_{1}k_{4}\ne0$.
The corresponding equations of motion are given as
\begin{subequations}
\begin{eqnarray}
\addtocounter{equation}{-1}
\label{isoeq2}
\addtocounter{equation}{1}
&&\hspace{-2cm}\ddot q_{1}=\frac{1}{r_{1}g_{2}} \bigg[k_{2}k_{4}(\Omega_{1}^2-\Omega_{2}^2)q_{1}^{1-r_1}q_{2}^{r_{2}}\nonumber\\
&&\hspace{-1cm}+q_{1}^{-1}\left\{(k_{1}k_{4}\Omega_{1}^{2}-k_{2}k_{3}\Omega_{2}^{2})q_{1}^{2}+(-k_{2}k_{3}+k_{1}k_{4})(-1+r_{1})r_{1}\dot q_{1}^{2}\right\} \bigg] ,\\
&&\hspace{-2cm}\ddot q_{2}=\frac{1}{r_{2}g_{2}} \bigg[-k_{1}k_{3}(\Omega_{1}^2-\Omega_{2}^2)q_{1}^{r_{1}}q_{2}^{1-r_2}\nonumber\\
&&\hspace{-1cm}+q_{2}^{-1}\left\{(-k_{2}k_{3}\Omega_{1}^{2}+k_{1}k_{4}\Omega_{2}^{2})q_{2}^{2}-(k_{2}k_{3}-k_{1}k_{4})(-1+r_{2})r_{2}\dot q_{2}^{2}\right\} \bigg].
\end{eqnarray}
\end{subequations}
The general solution of the above system of equations can be obtained by either explicitly integrating the Hamilton's equations of motion or by substituting the relation (\ref{harmq}) in (\ref{harmq1}) and solving the resultant algebraic equations for $q_1$ and $q_2$.
 The explicit general solution of (\ref{isoeq2}) can be given as
\begin{subequations}
\label{isosol2}
\begin{eqnarray}
&&q_1=\left(\frac{\frac{Bk_{2}}{\Omega_2}\sin(\Omega_2t+\delta_2)-\frac{Ak_{4}}{\Omega_1}\sin(\Omega_1t+\delta_1)}{k_{2}k_{3}-k_{1}k_{4}}\right)^{\frac{1}{r_{1}}}, (k_2k_3-k_1k_4)\ne 0,  \\
&& q_2=\left(\frac{\frac{Ak_{3}}{\Omega_1}\sin(\Omega_1t+\delta_1)-\frac{Bk_{1}}{\Omega_2}\sin(\Omega_2t+\delta_2)}{k_{2}k_{3}-k_{1}k_{4}}\right)^{\frac{1}{r_{2}}}.   
\end{eqnarray}
\end{subequations}
We note here that the above solution contains the required number (four) of arbitrary constants, namely $A,\,B,\,\delta_1,$ and $\delta_2$.  The obtained solution (\ref{isosol2}) is analytic and bounded and exhibits oscillatory behaviour for the choice $1/r_1$ and $1/r_2$ are positive integers.
\begin{figure} 
\centering 
\includegraphics[width=0.7\columnwidth]{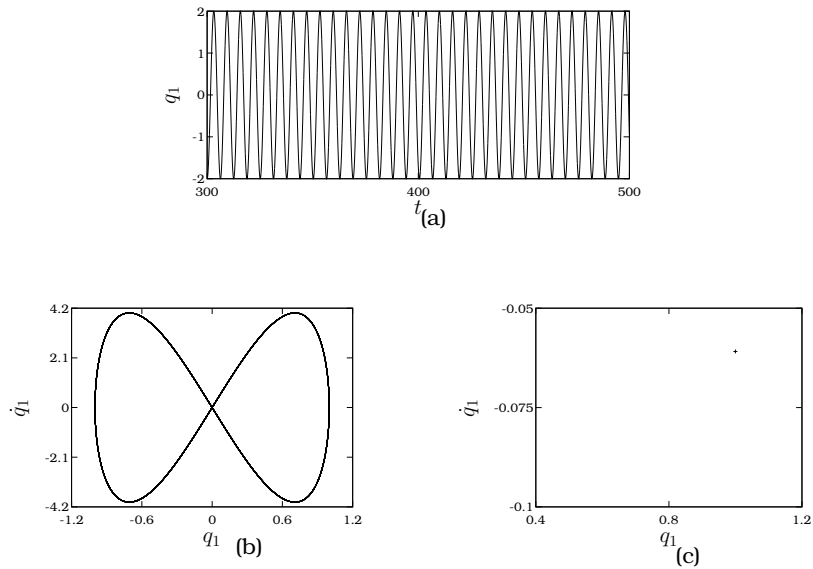} 
\caption{\label{fig3}Isochronous oscillations admitted by Eq. (\ref{isoeq2}) for the choice $\Omega_1:\Omega_2=1:1$, $r_{1}$= $r_{2}=\frac{1}{2}$ (a) Time series plot for $q_{1}$  (b) Projected $(q_1-\dot{q}_1)$ portrait (c) Poincar\'e section}.  
\end{figure} 

\begin{figure}
\centering 
\includegraphics[width=0.7\columnwidth]{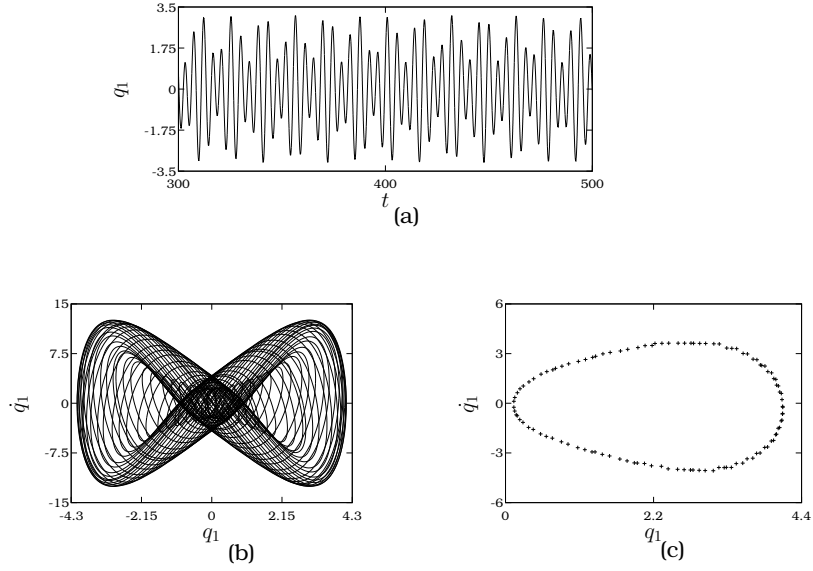}
\caption{\label{fig4}Quasiperiodic oscillations admitted by Eq. (\ref{isoeq2}) for the choice $\Omega_1:\Omega_2=1:\sqrt{2}$, $r_{1}$= $r_{2}=\frac{1}{2}$ (a) Time series plot for $q_{1}$ (b) Projected $(q_1-\dot{q}_1)$ portrait (c) Poincar\'e section } 
\end{figure} 

In Figs. 2 and 3, we plot the solution of Eq. (\ref{isoeq2}) for two different ratios of $\Omega_1$ and $\Omega_2$ with the parametric choice $r_{1}$= $r_{2}=\frac{1}{2}$,\, $k_{1}$=$k_{2}$=$k_{4}$=$1$ and $k_{3}$= $2$.  In Fig. 2, we choose $\Omega_1:\Omega_2=1:1$ which results in isochronous periodic oscillations.  In Fig. 3 we choose $\Omega_1:\Omega_2=1:\sqrt{2}$ which results in quasiperiodic oscillations with amplitude independent frequencies.

Having developed the method of constructing isochronous two dimensional oscillators, we extend the procedure to  $N$-dimensions in the next section.

\section{N-degrees of freedom generalization of Hamiltonians}

One can generalize the procedure of constructing isochronous Hamiltonian systems to $N$ degrees of freedom.  In order to do so, we have first extended the procedure to three dimensions and then generalized it to $N$ dimensions.  For the sake of brevity we do not present the results of three dimensional systems here.  
Proceeding further, to construct the N-dimensional isochronous Hamiltonians let us consider the following Hamiltonians
\begin{eqnarray}
H_{i}=\sum_{j=1}^{N}a_{ij}p_{j}\left(\frac{{\partial {Q_i(\underline{q})}}}{{\partial q_{j}}}\right)^{-1},\qquad i=1,2,\ldots,N\label{ndimham1}
\end{eqnarray}
which are canonically conjugate to the corresponding collective coordinates $Q_{i},\,\,\,i=1,2,\ldots N$. Here the functions $a_{ij}$'s are given by
\begin{eqnarray}
a_{ij}=\frac{Q_{ij}}{\Delta}, \nonumber
\end{eqnarray}
where
\begin{eqnarray}
\hspace{-1.6cm}Q_{ij}=Q_{iq_j}
\left|\begin{array}{cccccccc}
Q_{(i+1)q_{j+1}}&Q_{(i+1)q_{j+2}}&\ldots&Q_{(i+1)q_{N}}&Q_{(i+1)q_1}&Q_{(i+1)q_2}&\ldots&Q_{(i+1)q_{j-1}}\\
Q_{(i+2)q_{j+1}}&Q_{(i+2)q_{j+2}}&\ldots&Q_{(i+2)q_{N}}&Q_{(i+2)q_1}&Q_{(i+2)q_2}&\ldots&Q_{(i+2)q_{j-1}}\\
\vdots&\ddots&\ddots&\ddots&\ddots&\ddots&\ddots&\vdots\\
Q_{Nq_{j+1}}&Q_{Nq_{j+2}}&\ldots&Q_{Nq_N}&Q_{Nq_1}&Q_{Nq_2}&\ldots&Q_{Nq_{j-1}}
\end{array}
\right|,
\end{eqnarray}
\begin{eqnarray}
Q_{iq_j}=\left(\frac{\partial Q_i}{\partial q_j}\right),\qquad\Delta=\left|
\begin{array}{cccc}
Q_{1q_1}&Q_{1q_2}&\ldots&Q_{1q_N}\\
Q_{2q_1}&Q_{2q_2}&\ldots&Q_{2q_N}\\
\vdots&\ddots&\ddots&\vdots\\
Q_{Nq_1}&Q_{Nq_2}&\ldots&Q_{Nq_N}
\end{array}
\right|, \quad i,j=1,2,..N,             
\end{eqnarray}
 with $N+k=k$,  $k=0,1,2,\ldots,N-1$.  Here $Q_{ij}$ is an $N-1\times N-1$ determinant and $\Delta$ is an $N\times N$ determinant.
In order to construct the $N$-dimensional isochronous Hamiltonian system, let us substitute the $H_i$'s given by the relation (\ref{ndimham1}) into the Hamiltonian of the $N$-dimensional harmonic oscillator specified by
\begin{eqnarray}
\tilde{H}=\frac{1}{2}\sum_{i=1}^N\left( H_i^2+\Omega_i^2 Q_i^2\right),\label{ndimham2}
\end{eqnarray}
where $H_i$ and $Q_i$, $i=1,2,..N$, are the canonical momenta and coordinates, respectively.  Using this Hamiltonian $\tilde{H}$ and the corresponding Hamilton's equations of motion we find the following system of $N$ coupled second order ordinary differential equations (ODEs),
\begin{eqnarray}
\ddot{q}_i=\frac{1}{\Delta}\left(\sum_{k=1}^N\sum_{j=1}^N A_{ijk}\dot{q}_j\dot{q}_k+B_{i}\right),\quad i=1,2,\ldots,N,\,\,N>2\label{ndimsys}
\end{eqnarray}
where $A_{ijk}$ and $B_{ij}$, $j,k=1,2,\ldots,N$ are determinants of the form
\begin{eqnarray}
&&\hspace{-2.2cm}A_{ijk}=-\left|\begin{array}{cccc}
Q_{1q_{i+1}}&Q_{2q_{i+1}}&\ldots&Q_{Nq_{i+1}}\\
Q_{1q_{i+2}}&Q_{2q_{i+2}}&\ldots&Q_{Nq_{i+2}}\\
\vdots&\ddots&\ddots&\vdots\\
Q_{1q_{N}}&Q_{2q_{N}}&\ldots&Q_{Nq_{N}}\\
Q_{1q_{1}}&Q_{2q_{1}}&\ldots&Q_{Nq_{1}}\\
\vdots&\ddots&\ddots&\vdots\\
Q_{1q_{i-1}}&Q_{2q_{i-1}}&\ldots&Q_{Nq_{i-1}}\\
Q_{1q_{j}q_{k}}&Q_{2q_{j}q_{k}}&\ldots&Q_{Nq_{j}q_{k}}
\end{array}\right|_{N\times N},
\\\nonumber\\
&&\hspace{-2.2cm}
B_{i}=-\left|\begin{array}{cccc}
Q_1\Omega_1^2&Q_2\Omega_2^2&\ldots&Q_{N}\Omega_N^2\\
Q_{1q_{i+1}}&Q_{2q_{i+1}}&\ldots&Q_{Nq_{i+1}}\\
Q_{1q_{i+2}}&Q_{2q_{i+2}}&\ldots&Q_{Nq_{i+2}}\\
\vdots&\ddots&\ddots&\vdots\\
Q_{1q_{N}}&Q_{2q_{N}}&\ldots&Q_{Nq_{N}}\\
Q_{1q_{i-1}}&Q_{2q_{i-1}}&\ldots&Q_{Nq_{i-1}}
\end{array}\right|_{N\times N}.
\end{eqnarray}
The general solution of (\ref{ndimsys}) can be found after choosing the forms of $Q_i(q_1,q_2,\ldots,q_N)$ and following the procedure discussed for two coupled second order ODEs (\ref{isoeq2}).  Using the relation
\begin{eqnarray}
H_{i}=A_{i}\cos\bigg[\Omega_{i}t+\delta_{i}\bigg],    \nonumber
\end{eqnarray}
and 
\begin{eqnarray}
Q_{i}=\frac{A_{i}}{\Omega_{i}}\sin\bigg[\Omega_{i}t+\delta_{i}\bigg], \nonumber
\end{eqnarray}
where $A_{i}$ and $\delta_{i}$, $i=1,2,\ldots,N$, are arbitrary constants, or integrating the Hamilton's equations of motion corresponding to the Hamiltonian (\ref{ndimham2}), one can easily see that the canonical variables $q_{i},\,i=1,2,...,N$, also evolve periodically with a fixed period $T$ when $\Omega_i$'s are commensurate, for appropriate forms of $Q_{i}(q_{1},q_{2},\ldots,q_{N})$ such that the resultant solutions $q_i$s are analytic and single valued.

\section{Conclusion}
In this paper, we have shown that a class of singular Hamiltonian systems obtained through the $\Omega$-modified procedure are equivalent to constrained Newtonian systems. We have made use of the idea behind the procedure of $\Omega$-modified Hamiltonian and introduced suitable modifications to the procedure and developed a systematic procedure to construct a class of nonsingular $N$-dimensional isochronous systems exhibiting bounded isochronous oscillatory motion.   The Hamiltonian constructed by this procedure is nonsingular and the generated system is free from constraints and the Hamilton's equations and Newton's equation of motion are equivalent.  The procedure is first developed for the case of two-degrees of freedom systems and subsequently generalized to $N$-degrees of freedom systems.
The systems obtained through this procedure can also exhibit quasiperiodic oscillations with amplitude independent frequencies when the original linear harmonic oscillator frequencies are incommensurate.

\section*{Acknowledgments}
The work of VKC, RGP and ML is supported by DST--IRHPA research project. The work of  ML is also supported by a DAE Raja Ramanna Fellowship and a DST Ramanna Fellowship.

\section{Appendix}
The general solution of the system of two-coupled second order nonlinear ODEs (\ref{isoeq4}) can be found by the following procedure.  Using the point transformation 
\begin{eqnarray}
x=\sqrt{q_1},\quad y=\sqrt{q_2},
\end{eqnarray}  
we find that Eq. (\ref{isoeq4}) reduces to the following system of linear ODEs
\begin{eqnarray}
\ddot{x}=-\frac{\Omega^2}{2}(x+y),\quad
\ddot{y}=-\frac{\Omega^2}{2}(x+y).
\end{eqnarray}
 From the above system of ODEs, one can rewrite
\begin{eqnarray}
\ddot{x}+\ddot{y}=-\Omega^2(x+y),\quad
\ddot{x}-\ddot{y}=0.
\end{eqnarray}
Defining $z_1=x+y$ and $z_2=x-y$ we get
\begin{eqnarray}
\ddot{z}_1=-\Omega^2z_1,\qquad \ddot{z}_2=0,
\end{eqnarray}
whose general solutions are
\begin{eqnarray}
z_1=A\sin\Omega t+B\cos\Omega t,\quad z_2=I_1t+I_2,
\end{eqnarray}
where $A$, $B$, $I_{1}$, $I_{2}$ are integration constants.
Solving for $x$ and $y$ we get
\begin{eqnarray}
x=\frac{1}{2}\left(A\sin\Omega t+B\cos\Omega t+I_1t+I_2\right),\\
y=\frac{1}{2}\left(A\sin\Omega t+B\cos\Omega t-I_1t-I_2\right).
\end{eqnarray}
Redefining the above arbitrary integration constants as $A=\frac{H(0)}{\Omega}$, $B=\sqrt{q_1(0)}+\sqrt{q_2(0)}$, $I_{1}=C(0)$, $I_{2}=\sqrt{q_1(0)}-\sqrt{q_2(0)}$, we get the general solution of Eq. (\ref{isoeq4})
\begin{eqnarray}
q_1=\frac{1}{4}\left(2q_1(0)^2+\frac{H(0)\sin\Omega t}{\Omega}+Q(0)(\cos\Omega t-1)+C(0)t\right)^{2},\\
q_2=\frac{1}{4}\left(2q_2(0)^2+\frac{H(0)\sin\Omega t}{\Omega}+Q(0)(\cos\Omega t-1)-C(0)t\right)^{2},
\end{eqnarray}
where $q_1(0)$, $q_2(0)$, $H(0)$ and $C(0)$ are arbitrary constants and $Q(0)=\sqrt{q_1(0)}+\sqrt{q_2(0)}$.  The procedure can be straightforwardly extended to deduce the general solution of the $N$-coupled ODEs (\ref{isoeq1}).

Using the transformation $x_i=b_iq_i^{k_i}$ we find Eq. (\ref{isoeq1}) reduces to the form
\begin{eqnarray}
\ddot{x}_i=-\frac{\Omega^2}{N}(\sum_{m=1}^N x_m).
\end{eqnarray}
From the above system of ODEs, one can rewrite
\begin{eqnarray}
\sum_{m=1}^N\ddot{x}_m=-\Omega^2\sum_{m=1}^Nx_m,\qquad\ddot{x}_1-\ddot{x}_j=0,\quad j=2,3,\ldots,N
\end{eqnarray}
Defining the variables $z_{1}=\sum_{m=1}^N x_m$ and $z_{2j}=x_1-x_j,\quad j=2,3,\ldots,N$, we get
\begin{eqnarray}
\ddot{z}_{1}=-\Omega^2z_{1},\qquad\ddot{z}_{2i}=0.
\end{eqnarray}
Solving the above system of linear ODEs and rewriting the resultant solution in terms of the original variables $q_i$'s we get the general solution as
\begin{eqnarray}
q_n(t)=q_n(0)\left(1+\frac{a_n}{b_n}\frac{1}{(q_n(0))^{k_n}}\left(\frac{H(0)}{\Omega}\sin\Omega t+Q(0)(\cos\Omega t-1)\right)+C_n(0)t\right)^{\frac{1}{k_n}},\quad n=1,2,\ldots,N,
\end{eqnarray}
where $q_n(0)'s$, $\,H(0)$ and $C_n(0)$, $n=1,2,\ldots,N$ are integration constants fixed by the initial condition, $\sum_{n=1}^NC_n=0$ and $Q(0)=\sum_{m=1}^Nb_mq_m(0)^{k_m}$.

\end{document}